# Statistical Constraints on the Inflation Effective Potential from the *COBE DMR* Results

Hannu Kurki-Suonio

*Research Institute for Theoretical Physics, University of Helsinki, 00014 Helsinki, Finland*

Grant J. Mathews

*University of California, Lawrence Livermore National Laboratory, Livermore, California 94550*

(August 22, 1994)

We explore constraints on various forms for the effective potential during inflation based upon a statistical comparison between inflation-generated fluctuations in the cosmic microwave background temperature and the *COBE DMR* results. Fits to the first year $53A + B \times 90A + B$ cross correlation function using an effective potential of the form $V(\phi) = \lambda \phi^x / x!$, yield upper limits of $x \leq 107$ and $x \leq 397$ at the $1\sigma$ and $2\sigma$ confidence levels, while the same analysis produces a power law of index of $n = 1.10^{+0.47}_{-0.64}$ which would imply an underestimate of $x \leq 63$ from the simple analytic relation between $x$ and $n$. We also quantify new limits on the parameters for polynomial effective potentials. This work highlights the importance of a careful statistical treatment when seeking constraints on the inflation-generating effective potential.

Measurements of the large-scale anisotropy in the cosmic microwave background (CMB) by the Cosmic Background Explorer (*COBE*) Differential Microwave Radiometer (*DMR*) experiment have provided important support for the hot big bang model with inflation [1]. A favored explanation for the generation of the observed fluctuations in the CMB temperature is by the expansion of quantum fluctuations of a scalar field during the inflationary epoch [2–4]. The fact that the observed angular correlation function is more or less consistent with a scale-invariant Harrison-Zel'dovich spectrum of power on various angular scales is consistent with the predictions of inflationary models. Many papers, e.g. [5] – [12], have explored the basic relations between the shape of the effective potential during inflation and the observed power spectral index $n$ (i.e. $\langle |\delta_k|^2 \rangle \propto k^n$) for fluctuations on various scales. Although attempts have also been made to reconstruct the inflation-generating potential from the *COBE DMR* results [10–12], it is generally concluded [10] that it is not presently possible to obtain an accurate reconstruction of the potential.

The reason for this is that the spectral index is rather insensitive to the potential. For example, for an effective potential of the form $V(\phi) = \lambda \phi^x / x!$, it can be shown analytically [4] that

$$x = 120(1 - n) - 2. \quad (1)$$

(This result is for $x \ll 100$, where the slow-roll approximation is good). Thus, the limits on the spectral index obtained from fits [1] to the *COBE* angular correlation, $n = 1.1 \pm 0.5$, imply $x \lesssim 46$ at the one sigma level.

Most previous analyses have concentrated on the spectral index and the effects of the scalar and tensor contributions to the quadrupole moment (e.g. [5,8,9,11]). In this paper, however, we explicitly consider fits of various numerical inflation models to the *COBE DMR* cross correlation function. In this way confidence limits on various parameters for specific forms for the inflation generating potential can be better quantified and we include the specific parametric dependence of the contribution of the cosmic variance to the determination of levels of confidence.

We consider two simple forms for the effective potential,

$$V(\phi) = \frac{1}{x!} \lambda \phi^x \quad (2)$$

and

$$V(\phi) = \lambda \left( \tfrac{1}{8} \beta \phi^2 + \tfrac{1}{3} \alpha \phi^3 + \tfrac{1}{4} \phi^4 \right), \quad (3)$$

where we use Planck units, $c = m_{\rm Pl} = 1$.

The first potential form is often employed in "chaotic" inflation models [13], where $x$ is usually taken to have a value of 2 or 4. If $x > 4$, this is not a renormalizable potential, but could arise as an effective potential from another theory. The polynomial potential (3) represents the most general renormalizable potential with just one scalar field [14]. Although more complicated forms for the effective potential have been proposed (cf. [3,4]), this form is sufficiently general as it can represent the leading terms in an expansion of many different potentials. It has been discussed previously by Hodges et al. [14] in the context of generating non-Zel'dovich spectra over scales of galactic clustering.

Our goal is to compare the optimum values and projected confidence limits for the parameters $x$ (or $\alpha, \beta$) and $\lambda$ from specific inflation models with limits inferred from the uncertainty in the spectral index. The application of various statistical approaches for determining the spectral index and the ensemble-averaged quadrupole, $Q_{rms}$, has been discussed previously [15,16]. Some caveats regarding previous statistical analyses are summarized in [16]. This approach is contingent upon assuming a normal distribution of experimental errors and the cosmic variance. It has, however, been shown [17–19] that this is an adequate approximation when determining confidence limits.

The generation of the CMB anisotropy in inflationary models has been discussed extensively in the literature (e.g. [2–4,20]). The equations governing the evolution of the scalar field and the universal expansion are,

$$\ddot{\phi} + 3H\dot{\phi} + V'(\phi) = 0 \tag{4}$$

and

$$H^2 \equiv \left(\frac{\dot{R}}{R}\right)^2 = \frac{8\pi}{3}\left[V(\phi) + \frac{1}{2}\dot{\phi}^2\right]. \tag{5}$$

For simplicity, we assume instantaneous reheating as the scalar field approaches the minimum of the potential. Although this approximation introduces some uncertainty, it is the most optimistic scenario for determining the underlying potential. This approximation is valid as long as the universe experiences little inflation during reheating.

For the large scales observed by $COBE$ $DMR$, the CMB anisotropy is mostly due [21] to superhorizon fluctuations of the scalar field and metric tensor. Although clustering on subhorizon scales affects [20] the $COBE$ angular correlation function, for the present schematic study we restrict ourselves to the dominant superhorizon effects so that the amplitudes for a multipole expansion can be written

$$\langle a_l^2 \rangle = \langle a_l^2 \rangle_S + \langle a_l^2 \rangle_T \quad , \tag{6}$$

where the scalar contribution is

$$\langle a_l^2 \rangle_S = \frac{2l+1}{25\pi} \int_0^{\omega_{\max}} \frac{d\omega}{\omega} j_l(\omega)^2 \frac{H^4}{\dot{\phi}^2} \quad , \tag{7}$$

and the tensor contribution is

$$\langle a_l^2 \rangle_T = 36 l(l^2-1)(l+2)(2l+1) \int_0^{\omega_{\max}} d\omega\, \omega F_l(\omega)^2 H^2, \tag{8}$$

where from [23]

$$F_l(\omega) = \int_0^{s_{dec}} ds \left\{ -\frac{j_2[\omega(1-s)]}{\omega(1-s)} \right\}$$
$$\times \left[ \frac{2}{(2l-1)(2l+3)} j_l(\omega s) \right.$$
$$+ \frac{1}{(2l-1)(2l+1)} j_{l-2}(\omega s)$$
$$\left. + \frac{1}{(2l+1)(2l+3)} j_{l+2}(\omega s) \right] \quad , \tag{9}$$

with $\omega_{max} = 2(1+z_{eq})^{1/2}$ and $s_{dec} = 1 - (1+z_{dec})^{-1/2}$. The integrals in Eqs. (7-8) are over scales $\omega \equiv kr_0$, where $k$ is the comoving wavelength of the perturbation, and $r_0 = 2/H_0$ is the radius of the presently observable universe. We use $H_0 = 50$ km s$^{-1}$ Mpc$^{-1}$ and $z_{eq} = 10500$ and $z_{dec} = 1100$ for the redshifts of matter-radiation equality and hydrogen recombination, respectively. The quantities $H^4/\dot{\phi}^2$ and $H^2$ are evaluated at the epoch during inflation when the scale in question exits the horizon.

The use of $H^4/\dot{\phi}^2$ and $H^2$ in the integrands for the fluctuation amplitudes is based upon perturbation theory results obtained for exponentially expanding spacetimes, and is thus an approximation. However, the solution of Eqs. (4) and (5), as well as the integrals in (7), (8), and (9), were done numerically. We then convert from multipole expansion to the angular correlation function with the $COBE$ resolution [1],

$$C(\theta) = \frac{1}{4\pi} \sum_{l>2} \langle a_l^2 \rangle W_l^2 P_l(\cos\theta) \quad , \tag{10}$$

where we use the window function $W_l$ given in [22].

Since the amplitudes of the multipole moments arise from quantum fluctuations, they themselves are fluctuating quantities with a variance given by [23],

$$\langle \sigma_l^2 \rangle = \frac{2}{2l+1} \langle a_l^2 \rangle^2 \quad . \tag{11}$$

This defines the cosmic variance of the correlation function

$$\sigma(\theta)_{S,T}^2 = \frac{1}{(4\pi)^2} \sum_{l>2} \langle \sigma_l^2 \rangle_{S,T} \left[W_l^2 P_l(\cos\theta)\right]^2, \tag{12}$$

where the subscripts $S, T$ denote the scalar or tensor components.

When inflationary models generate significant amplitude for the higher multipoles, the cosmic variance diminishes due to the $(2l+1)^{-1}$ factor in (11). This occurs when the effective potential flattens during the descent.

The cosmic variance must be treated as a correlated error. We, therefore, utilize the generalization [24] of the $\chi^2$ distribution function

$$\chi^2 = \sum_i \sum_j \Delta C_i (C_{ij}^{-1}) \Delta C_j \quad , \tag{13}$$

where

$$\Delta C_i = C(\theta_i)_{obs} - C(\theta_i)_{calc} \quad , \tag{14}$$

and the covariance matrix $C_{ij}$ is given by,

$$C_{ij} = \sigma(\theta_i)_{obs}^2 \delta_{ij}$$
$$+ \left(\frac{T_{2.7}^2}{4\pi}\right)^2 \sum_l \frac{2}{2l+1} \left(\langle a_l^2\rangle_S + \langle a_l^2\rangle_T\right)^2$$
$$\times W_l^4 P_l(\cos(\theta_i)) P_l(\cos(\theta_j)) \quad , \tag{15}$$

where $\sigma(\theta_i)_{obs}$ is the $COBE$ experimental error bar and $T_{2.7}$ is the monopole temperature. Here we make use of

FIG. 2. Excluded regions of the $\alpha$ vs. $(\beta - \alpha^2)/\alpha^6$ plane. The light shading denotes the region excluded at the 68% confidence limit and the dark shaded region denotes the 95% confidence limit. The thin lines show contours of optimum $\lambda$ as labeled.

Significant deviations from a scale-invariant spectrum occur [14] as one approaches the line $\beta = \alpha^2$, $\alpha < 0$. We therefore present our results as a contour plot on the expanded $\alpha$ vs. $(\beta - \alpha^2)/\alpha^6$ plane in Fig. 2. The shaded regions on Fig. 2 identify excluded values at the $1\sigma$ (light shade) and $2\sigma$ (dark shade) level. The thin lines show contours of optimum values of $\lambda$ as labeled.

The reason for the excluded regions is a flattening of $V(\phi)$ to the right of the global minimum during the epoch that the small-scale ($< 10^\circ$) structure was formed. However, there was a step descent down the potential when the scales between 10° and 180° were formed. In this case, fluctuations on smaller scales dominate over the larger scales reducing the correlation function to zero at large angles. Just as important, however, is the fact that the cosmic variance for these fits is smaller due to the dominance of contributions from higher multipoles (smaller angular scales) with a smaller cosmic variance.

In summary, we have applied a statistical analysis directly to the inflation model rather than a naive analytic estimate from the uncertainty in the spectral index $n$. We found that one can only place rather weak constraints on the parameters of the effective potential. For a $\phi^x$ potential, one confounding factor is that a larger $x$ increases the cosmic variance. For a polynomial potential, one can have the opposite case, where a flattening of the potential increases the contribution of higher multipoles, reducing the cosmic variance. Terms which produce too much flattening of a polynomial effective potential can be excluded. The strongest constraints, however, are upon the overall amplitude $\lambda$ of the effective potential.

Stronger constraints on the inflation-generating effective potential will be possible as the statistics of the observed correlation function improve with more observing time. Indeed, we have made a preliminary application of the method described here to the second year $COBE$ data [25]. This reduces the upper limit on $x$. However, since the instrumental errors are now smaller, a more sophisticated analysis is warranted (e.g. [26]). Work along this line is currently in progress.

We acknowledge useful discussions with F. Graziani, G. F. Smoot, N. J. Snyderman, and M. S. Turner. Computations were carried out in part at the Center for Scientific Computing, Finland. HK-S thanks the Academy of Finland for financial support. Work performed in part under the auspices of the U. S. Department of Energy by the Lawrence Livermore National Laboratory under contract W-7405-ENG-48.